\documentclass[]{iopart}

\usepackage{iopams}

\usepackage[utf8]{inputenc}
\usepackage[T1]{fontenc}
\usepackage{graphicx}
\usepackage{epstopdf}

\begin{document}
\title{Two-field cosmological models and large-scale cosmic magnetic fields }

\author{Alexander A. Andrianov}
\address{Departament d'Estructura i Constituents de la Materia and
 Institut de Ci\`encies del Cosmos (ICCUB),\\
Universitat de Barcelona, 08028, Barcelona, Spain\\
V.A. Fock Department of Theoretical Physics, Saint Petersburg State University,
198504, S.~Petersburg, Russia
}
\author{Francesco Cannata}
\address{Istituto Nazionale di Fisica Nucleare, Via Irnerio 46, 40126 Bologna,
Italy}
\author{Alexander Y. Kamenshchik}
\address{Dipartimento di Fisica and Istituto Nazionale di Fisica Nucleare, Via Irnerio 46, 40126 Bologna,
Italy\\
L.D. Landau Institute for Theoretical Physics of the Russian
Academy of Sciences, Kosygin str. 2, 119334 Moscow, Russia }
\author{Daniele Regoli}
\address{Dipartimento di Fisica and Istituto Nazionale di Fisica Nucleare, Via Irnerio 46, 40126 Bologna,
Italy}
\eads{\mailto{andrianov@bo.infn.it}, \mailto{cannata@bo.infn.it}, \mailto{kamenshchik@bo.infn.it} and \mailto{regoli@bo.infn.it}}
\begin{abstract}
We consider two different toy cosmological models based on two fields (one normal scalar and one phantom) realizing the same evolution of the 
Bang-to-Rip type. One of the fields (pseudoscalar) interacts with the magnetic field breaking the conformal invariance of the latter.
The effects of the amplification of cosmic magnetic fields are studied and it is shown that the presence of such effects can discriminate between different cosmological models realizing the same global evolution of the universe. 
\end{abstract}


\maketitle

\section{Introduction}
The discovery of the cosmic acceleration \cite{cosmic} and the search for dark energy responsible for its 
origin \cite{dark} have stimulated the study of different field models driving the cosmological evolution.
Such a study usually is called the potential reconstruction \cite{recon}, because the most typical examples 
of these models are those with a scalar field, whose potential should be found to provide a given dynamics 
of the universe. In the flat Friedmann models with a single scalar field, the form of the potential and the time dependence 
of the scalar field are uniquely determined by the evolution of the Hubble variable (up to a shift of the scalar field). 
During last years the models with two scalar fields have also become very popular. This is connected with the fact that
some observations point out that the relation between the pressure and the energy density could be less than -1 \cite{obs}. 
Such equation of state arises if the matter is represented by a scalar field with a negative kinetic term. This field 
is called ``phantom'' \cite{phantom}.
Moreover, according to some observations \cite{obs1} the universe undergoes a transition between normal and phantom phase.
Such an effect is dubbed ``phantom divide line crossing'' \cite{divide}. 
In principle, the effect of phantom divide line crossing can be explained in the model with the only scalar field 
provided a special form of the potential and initial conditions is chosen \cite{AndCanKam} or in models with a non-minimally 
coupled scalar field \cite{non-minim}.
However, the models with two scalar fields, one standard and one phantom, look more ``natural'' for the description of the 
phantom divide line crossing \cite{two-field,two-field1,we-two-field}\footnote{These two fields   may have their origin in spontaneous breaking of
primordial symmetries of moduli space with complex potential \cite{two-field1}.}. 

In our preceding paper \cite {we-two-field} we have studied the procedure of reconstruction of the potential in two-field models. It was shown that there exists a huge variety of potentials and time dependences of the fields realizing the same cosmological evolution. Some concrete examples were considered, corresponding to the evolution beginning with the standard Big Bang singularity 
and ending in the Big Rip singularity \cite{Rip}.

One can ask oneself: what is the sense of  studying  different potentials and scalar field dynamics if they imply 
the same cosmological evolution?    The point is that the scalar and phantom field can interact with other  fields 
and   influence not only  the global cosmological evolution but also  other observable quantities.  

One of the possible effects of the presence of normal and phantom fields could be their influence on the dynamics of cosmic magnetic fields. The problem of the origin 
and of possible amplification of cosmic magnetic fields is widely discussed in the literature \cite{magnetic}. 
In particular, the origin of such fields can be attributed to primordial quantum fluctuations \cite{quantum} 
and their further evolution can be influenced by hypothetic interaction with pseudoscalar fields breaking 
the conformal invariance of the electromagnetic field \cite{break,Sorbo}. In the present paper we consider the evolution 
of magnetic fields created as a result of quantum fluctuations, undergoing the inflationary period 
with unbroken conformal invariance and beginning the interaction with pseudoscalar or pseudophantom fields 
after exiting the inflation and entering the Big Bang expansion stage, which is a part of the Bang-to-Rip scenario 
described in the preceding paper \cite{we-two-field}. We shall use different field realizations of this scenario and 
shall see how the dynamics of the field with negative parity influences the dynamics of cosmic magnetic fields. 
To our knowledge the possible influence of the two-field dynamics, (when one of two (pseudo)scalar fields is a phantom one) on the magnetic fields was not yet discussed in the literature.

Speaking of cosmic magentic fields we mean
the large-scale galactic, intergalactic or super-cluster magnetic
fields of order from $10^{-6} G$ to $10^{-11} G$ with correlation from 100 kpc to several
Mpc to the extent that they are originated from scalar and, possibly gauge field
fluctuations  after exiting the inflation. Their seeds may well
have $10^{-18} - 10^{-27} G$ or less (see \cite{magnetic}). 

The structure of the paper is as follows: in Sec.~2 we recall  the Bang-to-Rip scenario and describe some examples of  different dynamics 
of scalar and phantom fields; in Sec.~3 we introduce the interaction of the fields (phantom or normal) with an electromagnetic field 
and write down the corresponding equations of motion; in Sec.~4 we describe the numerical simulations of the evolution 
of magnetic fields and present the results of these simulations; Sec.~5 is devoted to concluding remarks.

\section{Cosmological evolution and (pseudo)-scalar fields}
We shall consider a spatially flat Friedmann universe with the metric
\begin{equation}
ds^2 = dt^2 - a^2(t) dl^2.
\label{Friedmann}
\end{equation}
Here the spatial distance element $dl$ refers to the so called comoving coordinates. The physical distance is 
obtained by multiplying $dl$ by the cosmological radius $a(t)$. 
We would like to consider the cosmological evolution characterized by the following time dependence of the 
Hubble variable $h(t) \equiv  \dot{a}/a$, where ``dot'' denotes the differentiation with respect to 
the cosmic time $t$:
\begin{equation}
h(t) = \frac{t_R}{3t(t_R-t)}.
\label{BtR}
\end{equation}
This scenario could be called ``Bang-to-Rip'' \cite{we-two-field} because it is obvious that at small values of $t$ the universe expands 
according to power law: $a \sim t^{1/3}$ while at $t \rightarrow t_R$ the Hubble variable explodes and one encounters 
the typical Big Rip type singularity. (The factor one third in (\ref{BtR}) was chosen for calculation simplicity). 
In our preceding paper \cite{we-two-field}) we considered a class of two-field cosmological models, where one field was 
a standard scalar field $\phi$, while the other was a phantom one $\xi$. The energy density of the system of these two 
interacting fields is 
\begin{equation}
\varepsilon = \frac12\dot{\phi}^2 - \frac12\dot{\xi}^2 + V(\phi,\xi).
\label{energy1}
\end{equation}
Analyzing the Friedmann equation \footnote{We use the following system of units $\hbar = 1, c =1$ and 
$8\pi G = 3$. In this system the Planck mass $m_P$, the Planck length $l_P$ and the Planck time $t_P$ are equal to $1$. Then when we need to make the transition to the ``normal'', say, cgs units, we should simply express the Planck units in terms of the cgs units.
In all that follows we tacitly assume that all our units are normalized by the proper Planck units. Thus, the scalar field entering 
as an argument into the dimensionless expressions should be divided by the factor $\sqrt{m_P/t_p}$.}

\begin{equation}
h^2 = \varepsilon,
\label{Friedmann1}
\end{equation}
we have shown that in contrast to models with one scalar field, there is huge variety of potentials $V(\phi,\xi)$ 
realizing a given evolution. Moreover, besides the freedom in the choice of the potential, one can choose different dynamics of the fields 
$\phi(t)$  and $\xi(t)$ realizing the given evolution. We have studied in \cite{we-two-field} some particular exactly 
solvable examples of forms of potentials and time dependences of fields $\phi$ and $\xi$ providing the evolution 
(\ref{BtR}). Here we shall present and apply some of them. 
Consider the potential~\footnote{The expression for the potential should be multiplied by the factor $m_P/(c l_P)$.}
\begin{equation}
V_{I}(\xi,\phi) = \frac{2}{9t_R^2}\cosh^6(-3\phi/4)\exp(3\sqrt{2}\xi).
\label{V-I}
\end{equation}
and the fields
\begin{equation}
\phi(t) = -\frac43 {\rm arctanh} \sqrt{\frac{t_R-t}{t_R}},
\label{phi1}
\end{equation}
\begin{equation}
\xi(t) = \frac{\sqrt{2}}{3} \ln \frac{t}{t_R-t}.
\label{xi1}
\end{equation}
(Here the expressions for the fields $\phi(t)$ and $\xi(t)$ should be multiplied by $\sqrt{m_P/t_P}$. For the relation between 
Planck units and cgs ones see e.g. \cite{Mukhanov-book}   
If we would like to substitute one of these  two fields by the pseudoscalar field, conserving the correct parity of 
the potential, we can choose only the field $\phi$ because the potential $V_{I}$ is even with respect to $\phi$, but not with respect to $\xi$. In what follows we shall call the model with the potential (\ref{V-I}), the pseudoscalar 
field (\ref{phi1})  and the scalar phantom (\ref{xi1}) ``model~$I$''.  

Consider another potential 
\begin{equation}
V_{II}(\xi,\phi) = \frac{2}{9t_R^2}\sinh^2(3\xi/4)\cosh^2(3\xi/4)\exp(-3\sqrt{2}\phi).
\label{V-II}
\end{equation}
with the fields 
\begin{equation}
\phi(t) = \frac{\sqrt{2}}{3} \ln \frac{t}{t_R-t},
\label{phi-new3}
\end{equation}
\begin{equation}
\xi(t) = \frac{4}{3} {\rm arctanh} \sqrt{\frac{t}{t_R}}.
\label{xi-new3}
\end{equation}
This potential is even with respect to  the field  $\xi$. Hence our model $II$ is based on the potential
(\ref{V-II}), the pseudophantom field (\ref{xi-new3}) and the scalar field (\ref{phi-new3}). 
They will be the fields with the negative parity which couple to the magnetic field.

\section{Post-inflationary evolution of a magnetic field  interacting with a pseudo-scalar or pseudo-phantom fields}
The action of an electromagnetic field interacting with a pseudoscalar or pseudophantom field $\phi$ is
\begin{equation}
S = -\frac14\int d^4x \sqrt{-g} ( F_{\mu\nu}F^{\mu\nu} + \alpha \phi F_{\mu\nu} \tilde{F}^{\mu\nu}),
\label{action} 
\end{equation}
where $\alpha$ is an interaction constant and the dual electromagnetic tensor $\tilde{F}^{\mu\nu}$ is defined as
\begin{equation}
\tilde{F}^{\mu\nu} \equiv \frac12 E^{\mu\nu\rho\sigma}F_{\rho\sigma},
\label{dual}
\end{equation}
where
\begin{equation}
E_{\mu\nu\rho\sigma} \equiv \sqrt{-g}\ \epsilon_{\mu\nu\rho\sigma},\ \ 
E^{\mu\nu\rho\sigma} \equiv -\frac{1}{\sqrt{-g}}\ \epsilon^{\mu\nu\rho\sigma}, 
\label{dual1}
\end{equation}
with the standard Levi-Civita symbol
\begin{equation}
\epsilon_{\mu\nu\rho\sigma} = \epsilon_{[\mu\nu\rho\sigma]},\ \epsilon_{0123} = +1.
\label{LC}
\end{equation}

Variating the action (\ref{action}) with respect to the field $A_{\mu}$ we obtain the field equations
\begin{equation}
\nabla_{\mu}F^{\mu\nu} = -\alpha\partial_{\mu}\phi\tilde{F}^{\mu\nu},
\label{Maxwell}
\end{equation}
\begin{equation}
\nabla_{\mu}\tilde{F}^{\mu\nu} = 0.
\label{Maxwell1}
\end{equation}
The Klein-Gordon equation for the pseudoscalar field is 

\begin{equation}
\nabla^{\mu}\nabla_{\mu}\phi + \frac{\partial V}{\partial \phi} = - \alpha F_{\mu\nu} \tilde{F}^{\mu\nu}.
\label{KG} 
\end{equation}
The Klein-Gordon equation for the pseudophantom field (which is the one that couples with the magnetic field in the model $II$) differs from equation (\ref{KG}) by change of sign in front of the kinetic term. 
In what follows we shall neglect the influence of magnetic fields on the cosmological evolution, i.e. we will discard the electromagnetic coupling in equation (\ref{KG}). 

If one wants to rewrite these formulae in terms of the three-dimensional quantities (i.e. the electric and magnetic fields) one can find the expression of the electromagnetic tensor in a generic curved background, starting from a locally flat reference frame --- in which it is well known the relation between electromagnetic fields and $F$ --- and using a coordinate transformation. It is easy to see that we have, for the metric (\ref{Friedmann}):
\begin{equation}
 F^{\mu\nu}=\frac{1}{a^2}\left(\begin{array}{cccc}0 & -aE_1 & -aE_2 & -aE_3\\
             aE_1 & 0 & B_3 & -B_2 \\
             aE_2 & -B_3 & 0 & B_1 \\
             aE_1 & B_2 & -B_1 & 0     
            \end{array}\right).
\end{equation}
The field equations (\ref{Maxwell}), (\ref{Maxwell1}) and (\ref{KG}) rewritten in terms of $\vec{E}$ and $\vec{B}$ become
\numparts\begin{equation}
\vec{\nabla}\cdot\vec{E} = - \alpha \vec{\nabla}\phi\cdot\vec{B}.
\label{Maxwell2}
\end{equation}
\begin{equation}
\partial_0(a^2\vec{E}) - \vec{\nabla}\times(a\vec{B}) = - \alpha[\partial_0\phi(a^2\vec{B}) - \vec{\nabla}\phi\times(a\vec{E})],
\label{Maxwell3}
\end{equation}
\begin{equation}
\partial_0(a^2\vec{B}) - \vec{\nabla}\times(a\vec{E}) = 0,
\label{Maxwell4}
\end{equation}
\begin{equation}
\vec{\nabla}\cdot\vec{B} = 0.
\label{Maxwell5}
\end{equation}\endnumparts

For a spatially homogeneous pseudoscalar field  equations (\ref{Maxwell2}) and (\ref{Maxwell3}) look like
\numparts\begin{equation}
\vec{\nabla}\cdot\vec{E} = 0
\label{Maxwell6}
\end{equation}
\begin{equation}
\partial_0(a^2\vec{E}) - \vec{\nabla}\times(a\vec{B}) = -\alpha\partial_0\phi(a^2\vec{B}).
\label{Maxwell7}
\end{equation}\endnumparts
Taking the curl of (\ref{Maxwell7}) and substituting into it 
the value of  $\vec{E}$ from (\ref{Maxwell4}) we obtain
\begin{equation}
\partial^2_0(a^2\vec{B})+h(t)\partial_0(a^2\vec{B})-\frac{\Delta^{(3)}(a^2\vec{B})}{a^2}-\frac{\alpha}{a}{\partial_0\phi\vec{\nabla}\times(a^2\vec{B})}=0,
\label{Maxwell8}
\end{equation}
where $\Delta^{(3)}$ stands for the three-dimensional Euclidean Laplacian operator.

Let us introduce 
\begin{equation}
\vec{F}(\vec{x},t) \equiv a^2(t)\vec{B}(\vec{x},t)
\label{F-def}
\end{equation}
and its Fourier transform
\begin{equation}
\vec{F}(\vec{k},t) = \frac{1}{(2\pi)^{3/2}} \int e^{-i\vec{k}\cdot\vec{x}} \vec{F}(\vec{x},t)d^3x.
\label{Fourier}
\end{equation}
Here the field $\vec{B}$ is an observable magnetic field entering into the expression for the Lorentz force.  
The field equation for  $\vec{F}(\vec{k},t)$ is 
\begin{equation}
 \ddot{\vec{F}}(\vec{k},t)+h(t)\dot{\vec{F}}(\vec{k},t)+\left[\left(\frac{k}{a}\right)^2-\frac{i\alpha}{a}\dot{\phi}\vec{k}
\times\right]
F(\vec{k},t)=0,
\label{Max-Four}
\end{equation}
where ``dot'' means the time derivative.
This last equation can be further simplified: assuming $\vec{k}=(k,0,0)$ and defining the functions $F_\pm\equiv (F_2\pm iF_3)/\sqrt{2}$ one arrives to
\begin{equation}
 \ddot{F}_{\pm} + h\dot{F}_{\pm}  +\left[\left(\frac{k}{a}\right)^2\pm\alpha\frac{k}{a}\dot{\phi}\right]F_{\pm}=0,
\label{Max-Four1}
\end{equation}
where we have omitted the arguments $k$ and $t$.  

Assuming that the electromagnetic field has a quantum origin (as all the fields in the cosmology of the early universe 
\cite{initial})  the modes of this field are represented by harmonic oscillators. Considering their vacuum 
fluctuations responsible for their birth we can neglect the small breakdown of the conformal symmetry and treat them as free.
In conformal coordinates $(\eta,\vec{x})$ such that the Friedmann metric has the form 
\begin{equation}
ds^2 = a^2(\eta)(d\eta^2 - dl^2),
\label{conformal}
\end{equation}
the electromagnetic potential $A_i$ with the gauge choice $A_0=0$, $\partial_jA^j=0$ satisfies the standard harmonic oscillator equation of motion
 \begin{equation}
 \ddot{A}_i + k^2 A_i = 0.
\label{harmonic}
 \end{equation}
Hence the initial amplitude of the field $A_i$ behaves as $A_i = 1/\sqrt{2k}$, while the initial amplitude of the 
 functions $F$ is $\sqrt{k/2}$. The evolution of the field $F$ during the inflationary period was described in 
\cite{Sorbo}, where it was shown that the growing solution at the end of inflation is amplified by some factor depending 
on the intensity of the interaction between the pseudoscalar field and magnetic field. 

Here we are interested in the evolution of the magnetic field interacting with the pseudoscalar field after inflation, 
where our hypothetic  Bang-to-Rip scenario takes place. More precisely, we would like to see how different types 
of scalar-pseudoscalar potentials and field dynamics providing the same cosmological evolution could be distinguished by their influence 
on the evolution of the magnetic field. We do not take into account the breaking of the conformal invariance during 
the inflationary stage 
and all the effects connected with this breakdown will be revealed only after the end of inflation and the beginning of 
the Bang-to-Rip evolution. This beginning is such that  the value of the Hubble parameter, 
characterizing this evolution is equal to that of the inflation, i.e. 
\begin{equation}
h(t_0) = \frac{t_R}{3t_0(t_R-t_0)} = h_{\rm inflation} \simeq 10^{33} {\rm s}^{-1}.
\label{beginning}
\end{equation}
In turn, this implies that we begin evolution at the time moment of the order of $10^{-33}{\rm s}$.

We shall consider both the components $F_{+}$ and $F_{-}$ and we shall dwell on the scenarios $I$ and $II$ 
described in the preceding section. 
Anyway, our assumption regarding the initial conditions for equation (\ref{Max-Four1}) can be easily modified in order to 
account for the previous possible amplification of primordial magnetic fields as was discussed by \cite{Sorbo}. 
Thus, all estimates for the numerical values of the magnetic fields in today's universe  should be multiplied by some 
factor corresponding to the amplification of the magnetic field during the inflationary stage.  Hence, our results refer more to 
differences between various models of a post inflationary evolution than of the real present values of magnetic fields, whose amplification might be also combined effect of different mechanism  \cite{magnetic,break,quantum,Sorbo}.

\section{Generation of magnetic fields: numerical results}

In this section we present the results of numerical simulations, for the two models $I$ and $II$ introduced in Sec.~2. 
In our models, the equation of motion for the modes $F_\pm(k,t)$ (\ref{Max-Four1}) reads\footnote{The reader can easily verify that this equation is obtained imposing the normalization $a|_{\rm today} = 1$, where today-time is taken to be near the crossing of the phantom divide line, i.e. at $t\simeq t_R/2$. This implies in turn that at the beginning of the ``Bang-to-Rip'' evolution the cosmological radius is $a(t_0) \simeq 10^{-17}$.}:
\begin{equation}
 \ddot{F}_{\pm} + \frac{t_R\ \dot{F}_{\pm}}{3t(t_R-t)} +
 \left[k^2\left(\frac{t_R-t}{t}\right)^{2/3}\pm\alpha\left(\frac{t_R-t}{t}\right)^{1/3} k\dot{\Phi}\right]F_{\pm}=0,
\label{edm}
\end{equation}
where $\Phi$ stands for the scalar field $\phi$ in the model~$I$ and for the phantom $\xi$ in the model~$II$, so that
\begin{equation}
\dot\Phi_I=\dot\phi= \frac23 \frac{\sqrt{t_R}}{t \sqrt{t_R - t}};\quad \dot\Phi_{II}=\dot\xi= \frac23 \frac{\sqrt{t_R}}{\sqrt{t} (t_R - t)}.
\label{dots}
\end{equation}

Equation (\ref{edm}) is solved for different values of the wave number $k$ and the coupling parameter $\alpha$.  
(The parameter $\alpha$ has the dimensionality inverse with respect to that of the scalar field; the wave number $k$ has the dimensionality of inverse length; the time $t_R = 10^{17}{\rm s}$).  Qualitatively we remark that in (\ref{Max-Four1}) the coupling term influence becomes negligible after some critical 
period. After that the magnetic fields in our different scenarios evolve as if the parameter $\alpha$ in  (\ref{Max-Four1}) 
had been put equal to zero. Indeed, it can be easily seen that the interaction term vanishes with the growth of the cosmological radius $a$. Then the distinction between the two models is to be searched in the early time behavior of the field evolution. 

Noting that in both our models $I$ and $II$ the time derivative $\dot{\Phi}$ is positive, by inspection of the linear term in equation (\ref{Max-Four1}) we expect the amplification 
to be mainly given for the mode $F_{-}$  provided the positive sign for $\alpha$ is chosen; so we will restrict our attention on $F_{-}$.
We can also argue that the relative strength of the last two terms in the left-hand side of (\ref{edm})  is crucial for determining the behavior of the solution: when the coupling term prevails (we remark that we are talking about $F_-$ so this term is \emph{negative} in our models) then we expect an amplification, while when the first term dominates we expect an oscillatory behavior.
For future reference it is convenient to define
\begin{equation}
{\bf A}(t;\Phi)\equiv \frac{k}{a(t)\alpha\dot{\Phi}}=\frac{k}{\alpha\dot\Phi}\left(\frac{t_R-t}{t}\right)^{1/3},
\label{A}
\end{equation}
which is just the ratio between the last two terms in the left-hand side of equation~(\ref{edm}).

Indeed our numerical simulations confirm these predictions.
Let us consider the model $I$ with $\alpha=1(t_P/m_P)^{1/2}$ and $k=10^{-55}l_P^{-1}$, where $l_P$ is the Planck length. Such a value of the wave 
number $k$ corresponds to the wave length of $1 kpc$ at the present moment.  
We obtain an early-time amplification of about 2 orders of magnitude, with the subsequent oscillatory decay. Notice that the parameter ${\bf A}$ in this model at the beginning is very 
small: this corresponds to the dominance of the term proportional to $\dot{\Phi}$ and, hence, to the amplification of the field 
$F_{-}$. At the time scale of the order of $10^{51} t_P$, where $t_P$ is the Planck time, this regime turns to that with big values of ${\bf A}$ where the 
influence of the term  proportional to $\dot{\Phi}$ is negligible.   

For the same choice of the parameters $\alpha$ and $k$ in the model $II$  
the amplification is absent. 

\begin{figure}[htp]\centering

\includegraphics{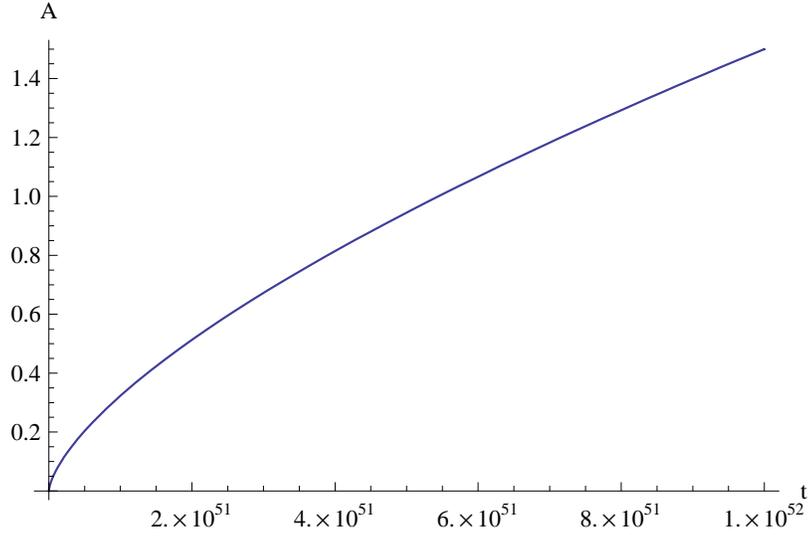}

\caption{Plot of the ratio ${\bf A}$ for the model $I$, with the parameter choice $\alpha=1 (t_P/m_P)^{1/2}$, $k=10^{-55}l_P^{-1}$. It can be easily seen that at a time scale of order $10^{51} t_P$ the ratio becomes greater than 1.}\label{mag1}
\end{figure}

\begin{figure}[htp]\centering

\includegraphics{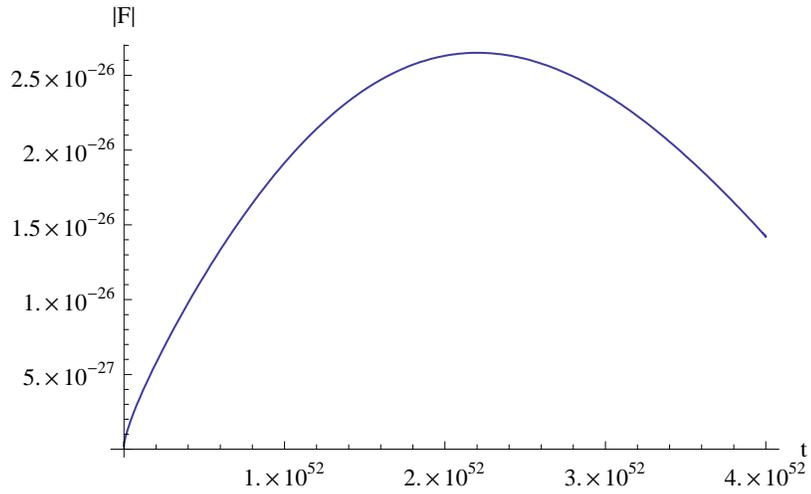}

\caption{Plot of the time evolution of the absolute value of the complex field $F$ (given in Planck units) in model $I$ with the parameter choice $\alpha=1 (t_P/m_P)^{1/2}$, $k=10^{-55}l_P^{-1}$. The behavior, as said above in the text, consists in an  amplification till a time of order $10^{51} t_P$, after which the oscillations begin.} \label{mag2}
\end{figure}

\begin{figure}[htp]\centering

\includegraphics{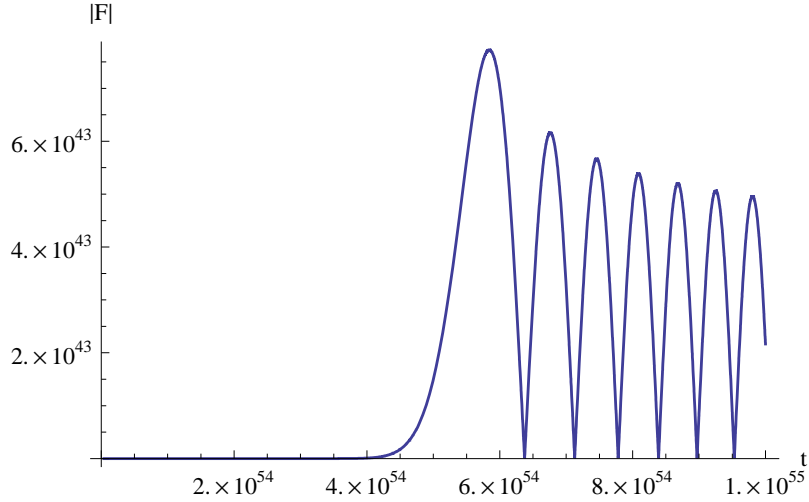}

\caption{Plot of the time evolution of the absolute value of the complex field $F$ (given in Planck units) in model $I$ with the parameter choice $\alpha=100 (t_P/m_P)^{1/2}$, $k=10^{-55}l_P^{-1}$. The behavior, consists in an  amplification till a time of order $10^{54} t_P$, after which the oscillations begin.} \label{mag3}
\end{figure}

In Figure~\ref{mag1}  we present the time dependence of the 
function ${\bf A}$ for the model $I$  for the values of $\alpha=1 (t_P/m_P)^{1/2}$ and $k=10^{-55}l_P^{-1}$ chosen above.
The Figure~\ref{mag2} manifests the amplification of the magnetic field in the model~$I$. 

Naturally the effect of amplification of the magnetic field grows with the  coupling constant $\alpha$ and diminishes when the  wave number $k$ increases. In Figure~\ref{mag3} we display the results for the case of $\alpha = 100 (t_P/m_P)^{1/2}$, which is admittedly extreme and possibly non realistic, but good for illustrative purposes. Here the 
amplification is more evident and extends for a longer time period. 
 
Let us try to make some estimates of the cosmic magnetic fields in the universe today, using the correlation functions.
The correlation function for the variable $F$ is defined as the quantum vacuum average  
\begin{equation}
G_{ij}(t,\vec{x}-\vec{y}) = \langle 0| F_i(t,\vec{x}) F_j(t,\vec{y})|0\rangle
\label{corr}
\end{equation}
and can be rewritten as 
\begin{equation}
G_{ij}(t,\vec{x}-\vec{y}) = \int \frac{d^3k}{(2\pi)^3} e^{-i\vec{k}\cdot (\vec{x}-\vec{y})} F_i(t,\vec{k})F_j^*(t,\vec{k}).
\label{corr1}
\end{equation}
Integrating over the angles, we come to 
\begin{equation}
G_{ij}(t,\vec{x}-\vec{y}) =  \int \frac{k^2 dk}{2\pi^2} \frac{\sin(k|\vec{x}-\vec{y}|)}{k|\vec{x}-\vec{y}|}F_i(t,k)F_j^*(t,k).
\label{corr2}
\end{equation}
To estimate the integral (\ref{corr2}) we notice that the main contribution to it comes from the region where $k \approx 1/|\vec{x}-\vec{y}|$ (see e.g.\cite{Mukhanov-book}) and it is of order
\begin{equation}
\frac{1}{L^3} |F_i(t,1/L)|^2,
\label{corr3}
\end{equation}
where $L = |\vec{x}-\vec{y}|$.
In this estimation the amplification factor is 
\begin{equation}
\left|\frac{F(1/L)}{\sqrt{1/L}}\right|,
\label{amplific}
\end{equation}
where the subscript $i$ is not present since we have taken the trace over polarizations.

Now we are in a position to give numerical values for the magnetic fields at different scales in the model $I$ for different values 
of the coupling parameter $\alpha$. These values (see Table~\ref{tab}) correspond to three values of the coupling parameter $\alpha$ (1,10 and 100 $(t_P/m_P)^{1/2}$) 
\footnote{It is useful to remark that at this length scales values of $\alpha$ less than 1 make the effect of the coupling of the magnetic field with the pseudoscalar field negligible.}
 and 
to two spatial scales $L$ determined by the values of the wave number $k$. We do not impose some physical restrictions on the value 
of $\alpha$. It is easy to see that the increase of $\alpha$ implies the growth of the value of the magnetic field $B$.  
We also shall consider a rather large value of the wave number $k$ corresponding to the physical wavelength $\sim 10^{-2}$ pc at the present moment when $a = 1$. While this scale 
looks too small for the description of  large-scale cosmic magentic fields, we use it for illustrative purposes to underline 
the strong dependence of the amplification of magnetic field on the corresponing wavelength values. 

Let us stress once again that we ignored the effects of amplification of the magnetic fields 
during inflation to focus on seizable effects during evolution. 

\begin{table}[h]
\caption{The Table displays the values of the magnetic field $B$ \\corresponding to the chosen values of $\alpha$ and $k$.
The length $L$\\ refers to the present moment when $a = 1$.}
\renewcommand{\arraystretch}{1.5}
\begin{tabular}{@{}lccc}
\br 
&$\alpha=1 \sqrt{t_P/m_P}$&$\alpha= 10 \sqrt{t_P/m_P}$ &$\alpha = 100 \sqrt{t_P/m_P}$\\ 
\mr
&&&\\
$k=10^{-55}l_P^{-1}\ (L=1kpc)$&$B \sim 10^{-67} G$&$B \sim	 10^{-60} G$&$B \sim 100 G$\\ \mr
&&&\\
$k=10^{-50}l_P^{-1}\ (L = 10^{-2}pc)$&$B\sim10^{-55} G$&$B \sim 10^{-49} G$&$B \sim 10^{13} G$\\ \br
\end{tabular}
\label{tab}
\end{table}

Finally notice that our quantum ``initial'' conditions correspond to physical magnetic fields which for presented values of $k$ are 
$B_{in} \sim k^2/a^2$ is equal to $10^{-34} G$ for $k = 10^{-55}l_P^{-1}$ and $B_{in} \sim 10^{-24} G$ for 
$k = 10^{-50}l_P^{-1}$.

\section{Conclusion}
We have seen that the evolution of the cosmic magnetic  fields interacting with a pseudoscalar (pseudophantom) field 
is quite sensitive to the concrete form of the dynamics of this field in two-field models
where different scalar field dynamics and potentials realize the same cosmological evolution. 

We confirm the sensitivity of the evolution of the magnetic field with respect to its helicity  
given the sign of the coupling constant $\alpha$ and that the $\Phi$ is a monotonic function of time 
(as it is really so in our models).
We give also some numerical estimates of the actual magnetic fields up to the factor of amplification 
of such fields during the inflationary period. The toy model of the Bang-to-Rip evolution studied in this paper,
cannot  be regarded as the only responsible for the amplification of cosmic magnetic fields 
implying their present observable values.
It rather complements some other mechanisms acting before. 
However, the difference  between cosmic magnetic fields arising  in various models 
(giving the same expansion law after the inflation) is essential. 
It may provide a discriminating test for such models.

Naturally, the influence of the interaction between a pseudoscalar (phantom)  field and a cosmic magnetic field 
on the dynamics of the latter depends on the velocity of change with time of the former. The larger is the time derivative 
of the pseudoscalar field, the more intensive is the growth of the magnetic field (remember that the evolution of the scalar field 
is  monotonic). The results of numerical calculations illustrated in Sec. 4 confirm these qualitative considerations.   
Moreover, one can see that there exists a certain range of values of the wave number $k$ (and hence of the corresponding wavelengths 
of the magnetic field) where the effect is stronger. Indeed, if $k$ is too small the interaction term is small as well and the evolution of the magnetic field is damped. On the other hand if the wave number $k$ is too large the interaction term is 
small compared to the oscillatory term, proportional as usual to $k^2$ and the evolution has practically oscillatory character. 
Thus, the study of interplay between the dynamics of global scalar fields providing the cosmological evolution and the magnetic fields looks promising. This interplay has another interesting aspect. The pseudoscalar-electromagnetic field interaction can imply a conversion of the photons into axions. Such an effect can cause the observable dimming of supenovae. 
While it was shown \cite{dimming} that this effect cannot mimic the cosmic acceleration, it could nevertheless mimic the dark energy fluid with a phantom equation of state \cite{dimming1}. Thus, the interrelation amongst an electromagnetic field, a scalar and a phantom field can reveal some surprises.

\ack
This work was partially supported by Grants RFBR  08-02-00923  and  LSS-4899.2008.2.
The work of
A.A. was  supported by
grants FPA2007-66665, 2005SGR00564, 2007PIV10046,  by the
Consolider-Ingenio 2010 Program CPAN (CSD2007-
00042) and Program RNP 2.1.1.1112.

\end{document}